\documentclass[
  journal=pasa,
  manuscript=research-paper, 
  year=2024,
  volume=37,
]{cup-journal}
\usepackage{graphicx}
\usepackage{dcolumn}
\usepackage{bm}
\usepackage{multirow}
\usepackage{threeparttable}
\usepackage[figuresright]{rotating}
\usepackage{caption}
\usepackage{scalerel}
\usepackage{subcaption}
\usepackage{hyperref}
\usepackage{float}
\usepackage{footnote}

\usepackage{microtype,siunitx,booktabs}
\usepackage{amsmath}
\usepackage{xcolor}

\hypersetup{colorlinks=true,citecolor=blue,linkcolor=blue,urlcolor=blue}

\title{A Multi-parameter Fuzzy Set Framework for Classifying Red, Blue, and Green Valley Galaxies}

\author{Amit Mondal}
\affiliation{Department of Physics, Visva-Bharati University, Santiniketan, 731235, West Bengal, India}
\email[Amit Mondal]{amitmondal.bwn95@gmail.com}

\author{Biswajit Pandey}
\affiliation{Department of Physics, Visva-Bharati University, Santiniketan, 731235, West Bengal, India}

\doi{xx.yyyy/pasa.zzzz.tt}

\received {dd Mmm YYYY}
\revised  {dd Mmm YYYY}
\accepted {dd Mmm YYYY}
\published{dd Mmm YYYY}


\begin{document}

\begin{abstract}
We present a data-driven fuzzy set theory framework for classifying galaxies into the red sequence, blue cloud, and green-valley populations using multiple observables from the Sloan Digital Sky Survey (SDSS DR18). Unlike traditional methods that rely on hard boundaries in colour or stellar mass space, our approach assigns continuous membership degrees based on sigmoidal functions derived from bimodal galaxy properties, including $(u-r)$ colour, specific star formation rate (sSFR), and $D4000$. The membership functions are constructed using Gaussian mixture modeling and combined through a conservative fuzzy minimum operator to obtain robust classifications. Applying this method to a volume-limited sample of $88\,255$ galaxies, we compare our results with the empirical classification of \citet{schawinski14}. We find that the fuzzy approach significantly reduces contamination in the red sequence and green-valley populations, yielding more physically consistent distributions in star formation activity and morphology. In particular, red galaxies exhibit a unimodal low-sSFR distribution, while green-valley galaxies show clearer signatures of morphological evolution. We further examine the dependence of active galactic nucleus (AGN) fraction on stellar mass and find no statistically significant differences between the two classification methods, indicating that global AGN trends are robust. However, clustering analysis reveals subtle but important differences: fuzzy-classified red and green-valley galaxies exhibit enhanced large-scale clustering, suggesting a stronger association with highly biased dark matter halos. These results demonstrate that fuzzy set-based classification provides a flexible, physically motivated, and robust alternative to traditional hard-cut methods. By avoiding arbitrary boundaries and leveraging multiple observables, our approach enables a more accurate and interpretable characterization of galaxy populations and their evolutionary pathways.
\end{abstract}

\section{Introduction}
\label{sec:intro}

The existence of the red sequence galaxies has been recognized for more than five decades \citep{baum59}. Over the intervening years, the broader structure of the galaxy colour distribution has gradually become clearer \citep{visvanathan81}, with a major advance driven by the availability of large, homogeneous galaxy samples from the Sloan Digital Sky Survey. Large optical galaxy surveys like SDSS \citep{york00, stout02} established that galaxies are not randomly distributed in optical colour vs absolute magnitude or in optical colour vs stellar mass space. They showed a bimodal distribution. These two modes represents two different populations one is red sequence and the other one is blue cloud, with a smaller population of galaxies lying in between them, which is commonly referred to as the green-valley \citep{wyder07}.

Several studies \citep{strateva01, blanton03, kauffmann03, baldry04b} have demonstrated that galaxies in the red sequence and the blue cloud possess distinctly different physical properties. Galaxies in the red sequence are typically massive and early type systems dominated by old stellar populations, mostly have a bulge dominated morphology such as ellipticals and lenticulars and having higher stellar mass \citep{kauffmann03, bell04, baldry04b}. These galaxies show little or no ongoing star formation and are therefore considered quiescent. Their red colours primarily arise from evolved, low-mass stars and the absence of young, massive stars. The blue cloud galaxies are generally late type, disk dominated systems such as spirals and irregulars and having lower stellar mass \citep{strateva01,kauffmann03, bell04, baldry04b}. They are actively forming stars, contain significant amounts of cold gas, and host young stellar populations, which give rise to their blue optical colours.
The correlation between morphology and colour is not absolute, with observations showing a significant number of elliptical galaxies in the blue cloud and spiral galaxies in the red sequence \citep{schawinski09,masters10}. 

The green-valley galaxies typically exhibit intermediate colours and moderate star formation rates and are widely interpreted as transitional systems evolving from the star forming blue population toward the quiescent red population \citep{martin07,schawinski14}. The existence of the green-valley suggests that galaxy evolution is not instantaneous but proceeds through a gradual or multi stage quenching process, possibly driven by mechanisms such as feedback from active galactic nuclei (AGN), environmental effects, or internal structural changes. A study by \citet{phillipps19} found that green-valley galaxies can undergo a gradual decline in star formation over $\sim2$--$4$ Gyr, potentially through gas depletion, without requiring an abrupt quenching event. In this context, recent studies using the Deep Extragalactic VIsible Legacy Survey (DEVILS; \citet{davies18}) provide further insight into the possible astrophysical origin of the transition from star-forming to quiescent galaxies through the green valley. \citet{davies25a} showed that galaxies occupying different regions of the sSFR--$M_\star$ plane exhibit distinct recent star-formation histories, indicating that their positions in this plane can correspond to different evolutionary pathways. \citet{davies25b} further showed that the physical processes associated with these pathways may vary across the stellar-mass scale corresponding to the minimum SFR dispersion. While stochastic star-formation processes appear to be important at lower stellar masses, rapidly declining star-formation histories at higher stellar masses are strongly associated with AGN activity, suggesting that AGN feedback may play an important role in the quenching of massive galaxies. These findings suggest that the evolution from star-forming to quiescent galaxies may be governed by different physical mechanisms across different stellar-mass regimes, with AGN feedback potentially contributing to the suppression of star formation and quenching at higher stellar masses. A detailed review article regarding green valley can be found in \citet{salim14}.

To understand the physical nature and evolutionary pathways of red, blue, and green galaxies is therefore central to modern galaxy evolution studies. Accurate classification of these populations is crucial for constraining quenching mechanisms, timescales, and the role of environment and internal processes in shaping the observed galaxy population.

In traditional method, galaxies have been classified into red or blue categories using hard empirical cuts in colour. Using SDSS galaxies, \citet{strateva01} introduced a colour cut at $(u - r) = 2.22$ to distinguish between blue cloud and red sequence populations.
\citet{baldry04a} distinguished between red and blue galaxies by modeling the observed $(u - r)$ colour distribution with a double Gaussian function.
Galaxy colour bimodality has been shown to vary systematically with luminosity, stellar mass, and environment \citep{balogh04, baldry06, pandey20a}. As a result, classifications based solely on colour can be inadequate, prompting the use of additional galaxy properties to distinguish the blue cloud from the red sequence.
 
The wide-field spectroscopic surveys most notably the Galaxy And Mass Assembly (GAMA) survey \citep{driver11} have significantly advanced our empirical understanding of colour bimodality and how it correlates with key galaxy properties. In particular, \citet{taylor15} demonstrated that the dominant uncertainty in measuring population-specific galaxy statistics arises from how red and blue classes are selected, arguing instead for modeling them as naturally overlapping populations in the colour--stellar mass plane. Further GAMA investigations have shown that colour bimodality systematically maps onto distinct morphological stellar mass functions \citep{baldry12} as well as mass--size relations and S{\'e}rsic indices \citep{lange15}.

In parallel, modern cosmological hydrodynamical simulations \citep[e.g.\ EAGLE,][IllustrisTNG, \citealp{nelson18}]{schaye15} and semi-analytic models have increasingly reproduced this bimodal architecture. \citet{bravo22} reconstructed the colour histories of GAMA galaxies alongside the SHARK semi-analytic model, applying Gaussian mixture modeling to show that galaxy populations are predominantly described by two evolving modes without requiring a distinct third population for the green valley. 

A wide range of studies have adopted this strategy by introducing empirical separation criteria in different parameter spaces, including the colour--magnitude diagram \citep{baldry04b, faber07, fritz14}, the colour--stellar mass plane \citep{taylor15}, and the colour--colour plane \citep{williams09, arnouts13, fritz14}.

A variety of methods have been proposed to identify the green-valley using different galaxy properties, resulting in definitions that are often imprecise and subjective, with criteria varying across studies. For example, \citet{schawinski14} define the green-valley using two empirical boundaries in the colour--stellar mass plane, while \citet{bremer18} classify red, blue, and green galaxies through three broad colour bins based on the surface density distribution in the colour--mass plane. \citet{coenda18} identify the green-valley in the (NUV$-r$) colour--stellar mass diagram using empirical lines and investigate the properties of transitional galaxies in different environments. In contrast, \citet{eales18} suggest that the green-valley does not constitute a distinct third population, but instead reflects a smooth transition from the blue cloud to the red sequence. An alternative approach based on spectral features was introduced by \citet{angthopo19}, who defined the green-valley using the 4000\,\AA\ break strength. This definition was later employed for a detailed analysis of the stellar populations of green-valley galaxies \citep{angthopo20}. \citet{quilley22} link galaxy morphology with evolutionary pathways and redefine the green-valley based on the mean colours of different Hubble types. Using a different approach, \citet{noirot22} employ the NUVrK colour--colour diagram to distinguish the blue cloud, green-valley, and red sequence. More recently, \citet{estrada23} identify the green-valley from the shape of the $\log(\mathrm{sSFR})$ distribution and investigate the morphological evolution of transitional galaxies in the CLEAR survey. In addition, \citet{brambila23} define the green-valley using empirical boundaries in the SFR--stellar mass plane and examine the influence of environment on the quenching of transitional systems. \citet{pandey23} introduced a parameter free approach to distinguish blue cloud and red sequence galaxies by applying Otsu’s image segmentation method \citep{otsu79}. Subsequently, \citet{pandey24} proposed a definition of the green valley in the colour-stellar mass plane based on entropic thresholding \citep{kapur85}.

Recently, there has been growing interest in identifying the green-valley galaxy population in a more robust manner, prompting the use of multiple galaxy observables to achieve a more reliable classification. \citet{nyiransengiyumva26} combined eight widely used criteria for selecting green-valley galaxies. They considered colour-based methods (e.g., $g-r$ and NUV$-r$), with the green-valley identified both visually and via Gaussian fitting, as well as SFR-based methods such as sSFR and SFR-$M_\ast$. 

Fuzzy set theory, introduced by \citet{zadeh65}, provides a flexible mathematical framework for handling vagueness and uncertainty in systems with overlapping or ill-defined boundaries. It has been widely applied in areas such as decision-making and automation \citep{zadeh73}, control systems \citep{lugli16, Lopatin18}, image processing and pattern recognition \citep{rosenfeld79, rosenfeld84, bezdek81}, and robotics \citep{wakileh88}.

In astronomy, it has been used for problems involving ambiguous classifications, including galaxy morphology \citep{spiekermann92}, star–galaxy separation \citep{mahonen00}, galaxy colour bimodality \citep{coppa11}, and colour-based galaxy classification that avoids sharp boundaries \citep{pandey20b}.

Despite significant progress, current classification schemes rely heavily on hard boundaries in parameter space, which fail to capture the intrinsic continuity of galaxy properties. This limitation is particularly severe in the green valley, where definitions remain inconsistent and often subjective across studies.

In this work, we introduce a multi-parameter fuzzy set theory framework that enables a continuous and physically motivated classification of galaxies. By incorporating multiple observables and constructing data-driven membership functions, our method avoids arbitrary thresholds and naturally captures transitional systems. This approach extends previous single-parameter fuzzy methods \citep{pandey20b} and provides a more robust characterization of galaxy populations.

The paper is organized as follows: \autoref{sec:datamethod} describes the simulation data and the methodology, \autoref{sec:results} presents the main results, and \autoref{sec:conclusion} summarizes the conclusions.

\section{Data and method of analysis}
\label{sec:datamethod}

The SDSS is one of the most extensive and influential redshift surveys conducted to date. Observations are carried out using a dedicated 2.5-meter telescope located at Apache Point Observatory in New Mexico. Over the years, SDSS has obtained photometric and spectroscopic measurements for millions of galaxies, stars, and quasars across a large fraction of the sky. In this work, we use data from SDSS Data Release 18 (DR18; \citealt{almeida}), the eighteenth public release of the survey.

The data were retrieved using \textit{Structured Query Language} (SQL) through the \textit{CasJobs} interface\footnote{\url{https://skyserver.sdss.org/casjobs/}}. We extracted galaxy-related spectroscopic and photometric information from the \textit{SpecObj}, \textit{PhotoObj}, and \textit{Photoz} tables. To ensure high-quality spectroscopic measurements, we selected only those galaxies with the \textit{scienceprimary} flag set to 1. We use observed galaxy colours without applying dust or k-corrections to preserve the intrinsic distribution of observables used for classification. Since our method relies on relative distributions rather than absolute values, this does not significantly affect the results.

Galaxy morphology is characterised using the concentration index, defined as $r_{90}/r_{50}$ \citep{shimasaku01}, where $r_{90}$ and $r_{50}$ denote the radii enclosing $90\%$ and $50\%$ of the Petrosian flux, respectively. These parameters are obtained from the \textit{PhotoObj} table. Additional galaxy properties, including stellar mass, specific star formation rate (sSFR), and metallicity, are derived from the \textit{StellarMassFSPSGranWideDust} table \citep{conroy09}. The sSFR measures the star formation rate normalised by the stellar mass of the galaxy, while metallicity represents the abundance of elements heavier than helium \citep{asplund09}. These quantities are estimated by comparing the observed photometric and spectroscopic properties of galaxies with predictions from stellar population synthesis models based on the Flexible Stellar Population Synthesis (FSPS) framework \citep{conroy09}. The SDSS spectra are obtained through fibres with a diameter of $3$ arcsec, which sample only a portion of the galaxy. Although aperture corrections can be applied to account for this limited coverage \citep{brinchmann04}, \cite{conroy09} show that estimates based solely on broadband photometry provide more reliable results. The strength of the $4000\,\mathring{A}$ break (D4000), which serves as an indicator of the mean stellar population age, is obtained from the \textit{galSpecIndx} table \citep{bruzual83, balogh99}.

Using SDSS DR18 data, we construct a volume-limited galaxy sample. We select a contiguous region of the sky bounded by right ascension $135^{\circ} \leq \alpha \leq 225^{\circ}$ and declination $0^{\circ} \leq \delta \leq 60^{\circ}$. The sample is defined by applying an extinction-corrected and $k$-corrected absolute magnitude cut in the $r$ band of $-23 \leq M_r \leq -21$. This magnitude range corresponds to a redshift interval of $0.0434 \leq z \leq 0.1175$. The chosen magnitude and redshift cuts ensure a volume-limited sample, minimizing selection biases and enabling a consistent comparison of galaxy populations. For the present analysis, we further restrict the range of several galaxy properties to remove extreme outliers. After applying all selection criteria, the final volume-limited sample contains $88\,255$ galaxies.



\begin{figure*}[h!]
\centering
    \includegraphics[width = \textwidth]{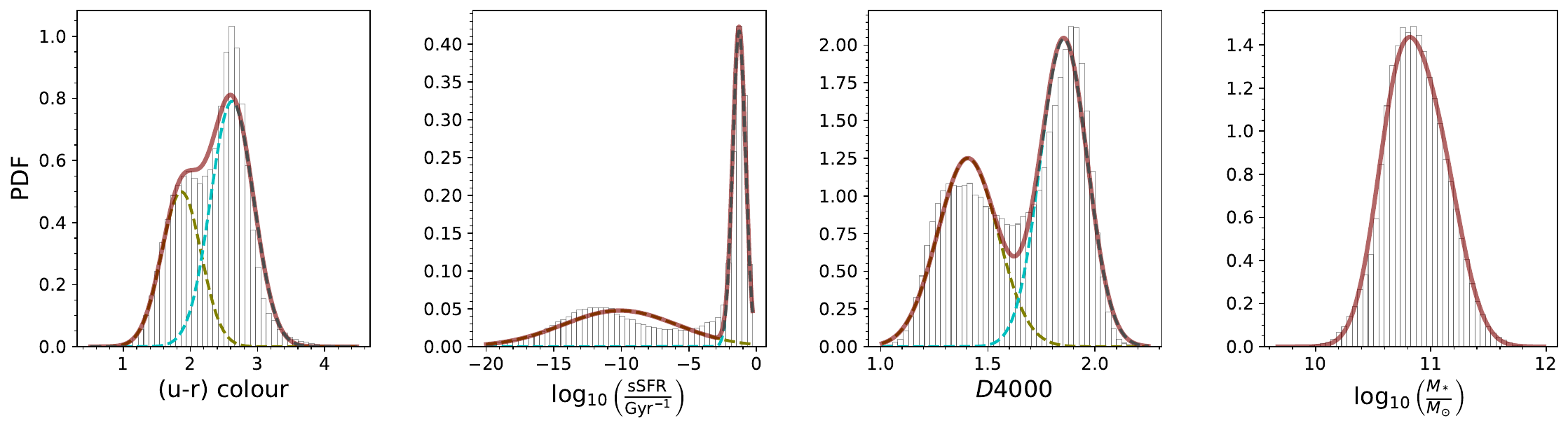}
    \caption{Probability density distributions of galaxy properties for a sample drawn from SDSS. The gray histograms show the data. The olive and cyan curves represent the two Gaussian components from the double-Gaussian fits, while the maroon curve shows their sum (the total model). From left to right, the panels correspond to the $(u-r)$ colour, $\log_{10}\left(\frac{sSFR}{{Gyr}^{-1}}\right)$, $D4000$, and $\log_{10}\left(\frac{M_*}{M_\odot}\right)$. }
 \label{fig:gmm}
\end{figure*}

\begin{figure*}[h!]
\centering
    \includegraphics[width = \textwidth]{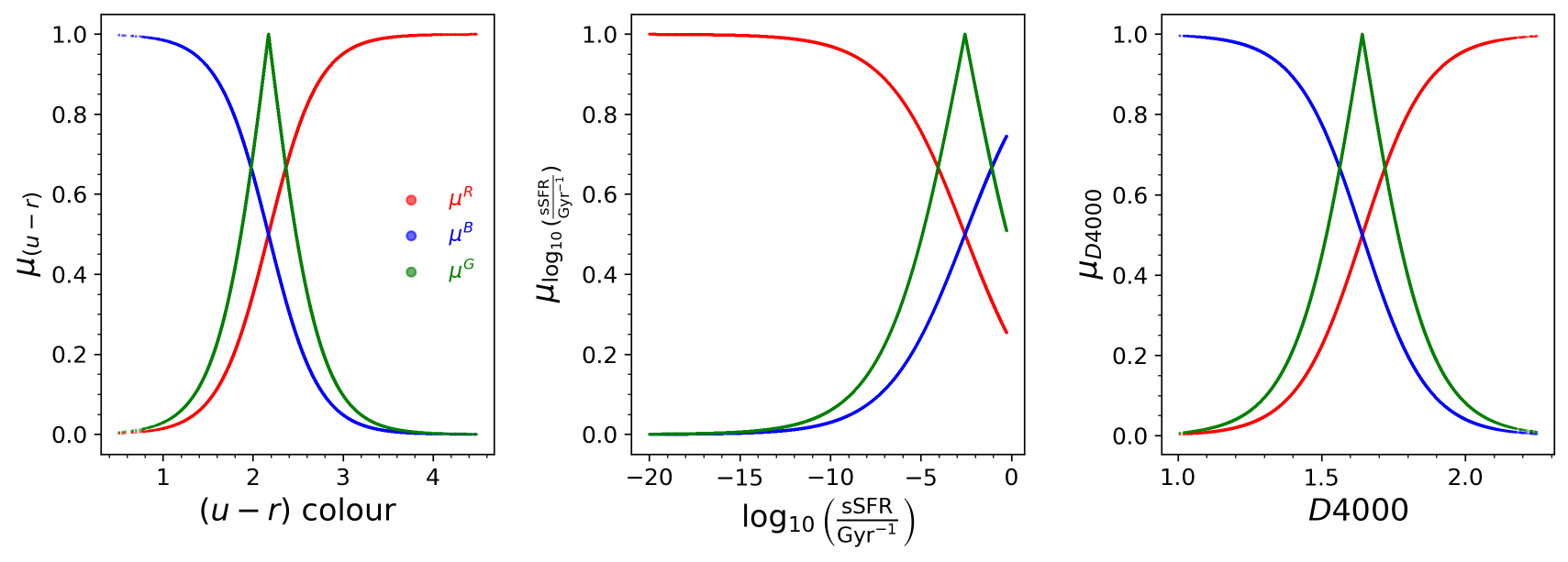}
    \caption{Membership functions derived for different galaxy properties. The panels show the redness (red), greenness (green), and blueness (blue) membership functions for each observable. From left to right, the properties are the $(u-r)$ colour, $\log_{10}\left(\frac{sSFR}{{Gyr}^{-1}}\right)$, and $D4000$.}
 \label{fig:memb_func}
\end{figure*}

\subsection{Fuzzy set and fuzzy membership functions}
\label{sec:fuzzy_set_memb}
Fuzzy set was first introduced by \citep{zadeh65}. Any fuzzy set $G$ within a universal set $X$ is defined by a membership function $\mu_{\scaleto{G}{3.5pt}}$, can be represented as,
\begin{equation}
    G = \{(x, \mu_{\scaleto{G}{3.5pt}}(x)) \mid x \in X\}.
\end{equation}
The value of $\mu_{\scaleto{G}{3.5pt}}(x)$ ranges between 0 to 1. $\mu_{\scaleto{G}{3.5pt}}(x) = 1$ denotes full membership, $\mu_{\scaleto{G}{3.5pt}}(x) = 0$ indicates non-membership, and intermediate values $(0 < \mu_{\scaleto{G}{3.5pt}}(x) <1)$ correspond to partial membership. This allows fuzzy sets to model vagueness and uncertainty effectively. The membership function of a fuzzy set can take various forms, such as triangular, trapezoidal, Gaussian, or sigmoidal, which depends on the specific context or application.

Traditional hard or crisp classification schemes cannot adequately capture the gradual transitions that exist between two populations. The boundary region between any two classes is particularly problematic, as hard cut classifications often become contaminated. This contamination arises because strict cuts may exclude genuine members, while relaxed cuts may include objects that do not truly belong to the population. These limitations motivate the use of fuzzy sets in classification schemes, as they allow for a more realistic treatment of transitional regions and overlapping populations.

The fuzzy approach allows each galaxy to simultaneously belong to red, blue, and green populations with varying degrees, rather than forcing a binary assignment. This reflects the continuous nature of galaxy evolution and avoids misclassification near population boundaries. Thus, we do not classify galaxies as purely red, blue, or green, instead we assign each galaxy a degree of redness, blueness, or greenness. Using SDSS data, we define fuzzy sets corresponding to these three colour categories. There are several galaxy properties that show clear bimodality, exhibiting distinct and well separated behaviour between red and blue galaxy populations. We include those properties in our analysis to construct fuzzy sets of redness of galaxies using each of the properties. The fuzzy set corresponding to `redness' of galaxies using a property p can be defined as,
\begin{equation}
 R_p = \{p,\, \mu_{\scaleto{p}{3.5pt}}^{\scaleto{R}{3.5pt}}(p) \mid p \in P\},    
\end{equation}
where, $P$ denote the universal set corresponding to the stellar property p of all galaxies. We adopt sigmoidal membership functions to capture the gradual transition between the two bimodal populations, red and blue galaxies. This functional form ensures a smooth transition between the two populations and naturally represents intermediate systems. The sigmoidal membership function is given by
\begin{equation}
\mu_{\scaleto{p}{3.5pt}}^{\scaleto{R}{3.5pt}}(p)=\frac{1}{1+e^{\pm a(p-c)}}.
\label{eq:red_memb_p}
\end{equation}
Here, $a$ is the steepness, and $c$ is the midpoint ($\mu_{\scaleto{p}{3.5pt}}^{\scaleto{R}{3.5pt}}(p) = 0.5$) of the sigmoid curve. A `$-$' sign signifies that higher $p$ corresponds to more redder galaxies, while a `$+$' sign denotes the exact opposite trend. Any galaxy has higher redness implies having smaller blueness. Therefore, redness and blueness are complementary. We can define the blueness fuzzy set using the $p$ is $B_p$, which is the fuzzy compliment of $R_p$. 
The membership function corresponding to the blueness fuzzy set can be defined as,
\begin{equation}
\mu_{\scaleto{p}{3.5pt}}^{\scaleto{B}{3.5pt}}(p) = 1 - \frac{1}{1+e^{\pm a(p-c)}}. 
\label{eq:blue_memb_p}
\end{equation}
The fuzzy set corresponding to the greenness of galaxies is the fuzzy intersection of redness and blueness fuzzy sets, i.e. $G_p = R_p \cap B_p$. The membership function corresponding to it is,
\begin{equation}
\mu_{\scaleto{p}{3.5pt}}^{\scaleto{G}{3.5pt}}(p) = 2 \min \Big\{ \mu_{\scaleto{p}{3.5pt}}^{\scaleto{R}{3.5pt}}(p), \,\mu_{\scaleto{p}{3.5pt}}^{\scaleto{B}{3.5pt}}(p) \Big\}.
\label{eq:green_memb_p}
\end{equation}
The factor 2 is used for normalization \citep{pandey20b}. To obtain the overall redness membership when multiple galaxy properties are used, we combine all individual redness membership values using the fuzzy minimum operator, 
\begin{equation}
\mu_{\scaleto{R}{3.5pt}}(p) =\min \Big\{ \mu^{\scaleto{R}{3.5pt}}_{\scaleto{p_1}{3.5pt}}(p_1), \,\mu^{\scaleto{R}{3.5pt}}_{\scaleto{p_2}{3.5pt}}(p_2), \dots \,\mu^{\scaleto{R}{3.5pt}}_{\scaleto{p_n}{3.5pt}}(p_n) \Big\}.
\label{eq:red_memb}
\end{equation}
Here, $p_1$, $p_2$,... $p_n$ denote different galaxy properties selected for the fuzzy classification of galaxies. 
We adopt the fuzzy minimum operator to combine membership functions across different observables. This conservative choice ensures that a galaxy is classified as red only if all relevant properties consistently support that classification. In other words, by adopting the minimum operator, the overall redness membership is limited by the least supportive property, ensuring that no single favorable attribute can dominate the classification. This deliberately restrictive approach avoids overestimating redness membership when any property deviates from redness behavior. 
The corresponding blueness membership is then defined by the complement of the combined redness criteria, effectively representing how strongly a galaxy departs from red characteristics and providing a continuous description of the transition. 
The overall blueness membership functions can be written as,
\begin{equation}
\mu_{\scaleto{B}{3.5pt}}(p) = 1 - \mu_{\scaleto{R}{3.5pt}}(p).
\label{eq:blue_memb}
\end{equation}
And the overall greenness membership functions is, 
\begin{equation}
\mu_{\scaleto{G}{3.5pt}}(p) = 2 \min \Big\{\mu_{\scaleto{R}{3.5pt}}(p), \,\mu_{\scaleto{B}{3.5pt}}(p) \Big\}.
\label{eq:green_memb}
\end{equation}

It is important to emphasize that membership functions do not represent the probability of a galaxy being `red', `blue', or `green'. Instead, they quantify the degree of possibility that a galaxy belongs to a given fuzzy set. Unlike likelihood functions in probability theory, membership functions are not required to be normalise to unity. Fuzzy set theory is therefore based on possibility rather than probability. Although both frameworks describe uncertainty, they differ fundamentally.

\section{Results and Discussions}
\label{sec:results}
\subsection{Classifications of red, blue, and green galaxies in SDSS data}

We apply our fuzzy galaxy classification technique to separate red, blue, and green galaxies from the sample constructed using SDSS data. The classification is based on three bimodal galaxy observables: $(u - r)$ colour, $\log_{10}(\mathrm{sSFR}/\mathrm{Gyr}^{-1})$, and $D4000$. As shown in \autoref{fig:gmm}, these parameters exhibit clear bimodal distributions, making them well suited for constructing fuzzy membership functions. In contrast, stellar mass does not display strong bimodality and is therefore excluded from the membership construction.

For each observable, we derive sigmoidal membership functions representing redness, blueness, and greenness (see \autoref{fig:memb_func}). To estimate the membership function for each property, the parameters $a$ and $c$ must be determined. For the properties $(u-r)$ colour and $D4000$ used in our analysis, larger values correspond to galaxies being in the redder regime. Consequently, in \autoref{eq:red_memb_p}, the `$-$' sign appears before the parameter $a$ in place of the `$\pm$' sign. On the other hand, for the property $\log_{10}\left(\frac{sSFR}{\mathrm{Gyr}^{-1}}\right)$, smaller values correspond to galaxies being in the redder regime. Therefore, in \autoref{eq:red_memb_p}, the `$+$' sign appears before the parameter $a$ instead of the `$\pm$' sign.
\\
We model the distribution of these three properties using a two-component Gaussian mixture with the \texttt{scikit-learn} package in Python \citep{pedregosa11}. The GMM fitted properties i.e. $(u-r)$ colour, $\log_{10}\left(\frac{sSFR}{{Gyr}^{-1}}\right)$, and $D4000$ are shown in the \autoref{fig:gmm}.

\begin{table}[t]
\centering
\caption{Parameters of the sigmoidal membership function for different galaxy observables.}
\begin{tabular*}{\columnwidth}{@{\extracolsep{\fill}}lcc}
\hline
Observable & $a$  & $c$  \\
\hline
$(u-r)$ colour & 3.584 & 2.171 \\
$\log_{10}\left(\frac{sSFR}{{Gyr}^{-1}}\right)$  & 0.468 & -2.580 \\
$D4000$ & 8.790 & 1.641 \\
\hline
\end{tabular*}
\label{tab:sigmoid_parameters}
\end{table}

We employ a two-component Gaussian mixture model with ten independent initializations, using the default k-means-based initialization without imposing any prior on the Gaussian parameters. To assess the sensitivity of the fitting procedure, we also consider two alternative approaches: initializing the Gaussian means near the observed distribution peaks and fixing the means at these positions. Although peak initialization yields a lower Bayesian Information Criterion (BIC), the resulting intersection points differ only slightly from those obtained using the original approach. The fixed-mean approach produces similarly small changes in the intersection points. We therefore retain the original unconstrained Gaussian mixture model, as the estimated transition parameters are relatively insensitive to these alternative fitting assumptions.

The parameter $c$ is defined as the intersection point of the two Gaussian components near their respective means. In traditional hard-cut classification schemes, this point serves as the boundary separating red and blue galaxies. Physically, this corresponds to the point of maximum confusion, where the classification uncertainty is highest. Accordingly, the membership degree of the fuzzy set representing red galaxies equals 0.5 at this point for any galaxy property.
The value of the parameter $c$ for the three galaxy observables $(u-r)$ colour, $\log_{10}\left(\frac{sSFR}{{Gyr}^{-1}}\right)$, and $D4000$ are tabulated in \autoref{tab:sigmoid_parameters}. \\

The next step is to determine the steepness parameter $a$ of the sigmoidal curve. The value of $a$ controls how gradual the transition is between red and blue galaxies and primarily depends on the standard deviations of the two Gaussian components. To estimate the steepness of the sigmoid, we use the width of its central transition region, defined as the interval over which the function increases from $10\%$ to $90\%$ of its maximum value. Based on this definition, we obtain
\begin{equation}
a = \frac{2\ln(9)}{2\big(\sigma_1 + \sigma_2\big)},
\end{equation}
where $\sigma_1$ and $\sigma_2$ are the standard deviations of the two Gaussian components associated with any galaxy property.
The value of the parameter $a$ for the three galaxy observables $(u-r)$ colour, $\log_{10}\left(\frac{sSFR}{{Gyr}^{-1}}\right)$, and $D4000$ are tabulated in \autoref{tab:sigmoid_parameters}. \\
Thus, we determine the membership functions for the fuzzy set of red galaxies in a data-driven manner using each galaxy property via \autoref{eq:red_memb_p}. The corresponding membership functions for blue galaxies are obtained from \autoref{eq:blue_memb_p} while those for green galaxies are derived using \autoref{eq:green_memb_p}. The variation of the membership functions of red, blue, and green galaxies, based on all three galaxy properties as a function of the associated property, is shown in \autoref{fig:memb_func}.

In all three panels of \autoref{fig:memb_func}, the red curve represents the redness membership function, the blue curve represents the blueness membership function, and the green curve represents the greenness membership function, corresponding respectively to the galaxy properties $(u-r)$ colour, $\log_{10}\left(\frac{sSFR}{{Gyr}^{-1}}\right)$, and $D4000$.

Finally, we compute the combined redness membership function by combining the redness membership functions of all three galaxy properties using \autoref{eq:red_memb}. Similarly, the combined blueness and greenness membership functions are obtained using \autoref{eq:blue_memb} and \autoref{eq:green_memb}, respectively. These combined membership functions are the identifires of the classification of the red, blue, and green classes of galaxies. 

Galaxies with a combined redness membership function $\mu_{\scaleto{R}{3.5pt}} > 0.5$ are classified as red sequence, those with $\mu_{\scaleto{B}{3.5pt}} > 0.5$ as blue cloud, and those with $\mu_{\scaleto{G}{3.5pt}} > 0.5$ as green-valley. The classification threshold of 0.5 is chosen as a natural decision boundary corresponding to maximum ambiguity in membership. The choice ensures that galaxies are classified based on their dominant membership.

\begin{figure*}[h!]
\centering
    \includegraphics[width = \textwidth]{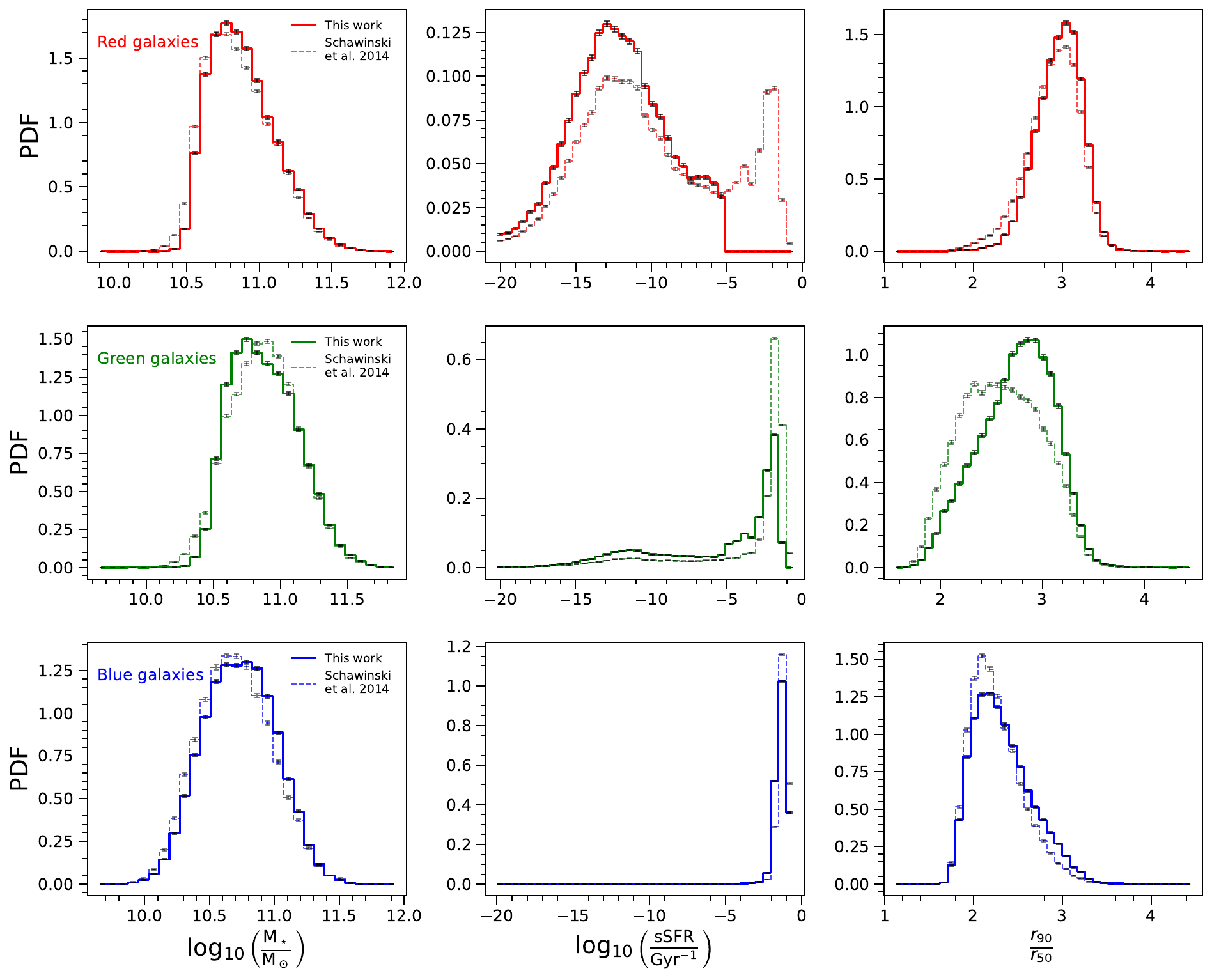}

    \caption{The left panels of the figure shows the stellar mass distributions, middle panels shows the sSFR distributions, and the right panels shows the concentration index distributions red-sequence, green-valley, and blue-cloud galaxies (from top to bottom), identified using the fuzzy based classification (solid line) method and the method proposed by \citet{schawinski14} (dashed line). The $1\sigma$ error bars are estimated from 50 Monte Carlo realizations of the observed quantities, incorporating the measurement uncertainties and their propagation.}
 \label{fig:rgb_enq}
\end{figure*}

\begin{figure*}[h!]
\centering
    \includegraphics[width = \textwidth]{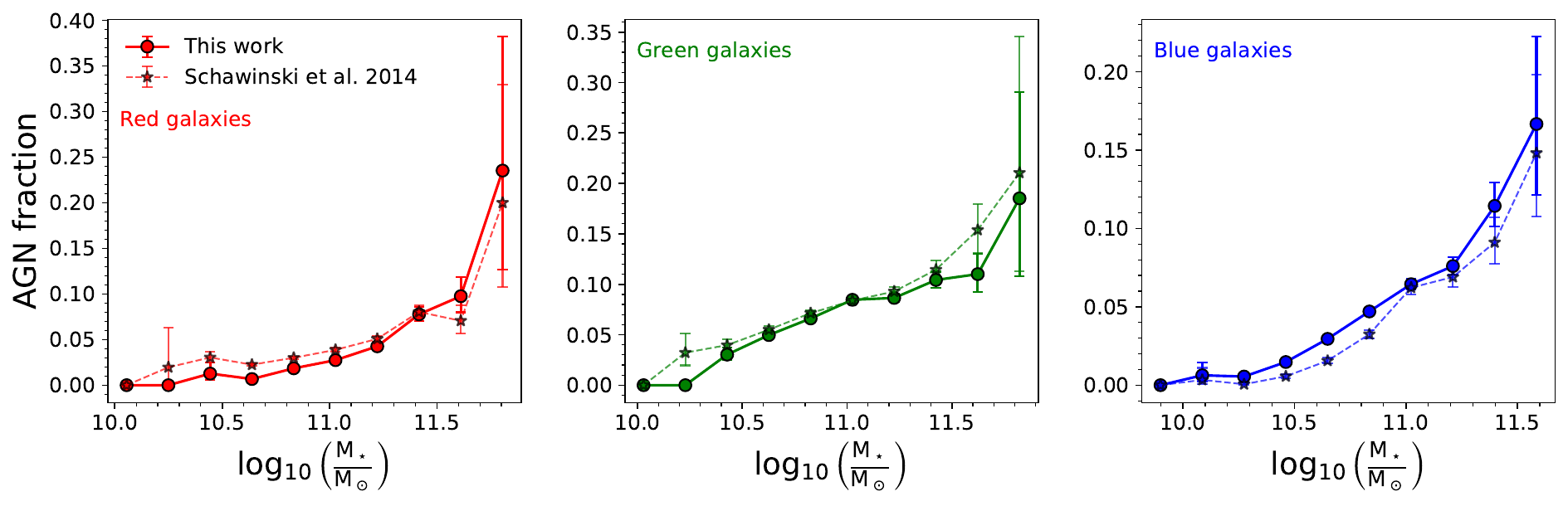}

    \caption{ The different panels of the figure show the AGN fraction as a function of stellar mass for red-sequence, green-valley, and blue-cloud galaxies (from left to right), identified using the fuzzy-based classification method (solid lines) and the method proposed by \citet{schawinski14} (dashed lines). Errorbars show 1$\sigma$ uncertainties using beta distribution quantile technique.}
 \label{fig:rgb_agn}
\end{figure*}

To validate our fuzzy classification framework, we perform a comparison with the empirical method proposed by \citet{schawinski14}. They proposed an empirical method to classify galaxies into red, blue, and green populations based on their location in the colour-stellar mass diagram. In this scheme, green-valley galaxies are defined as those lying between two straight lines in the $(u-r)$ colour versus stellar mass plane. The equations of these boundary lines are

\begin{equation}
u-r = -0.24 + 0.25\,\log_{10}\!\left(\frac{M_\ast}{M_\odot}\right)
\label{eq:stline_1}
\end{equation}  

\begin{equation}
u-r = -0.75 + 0.25\,\log_{10}\!\left(\frac{M_\ast}{M_\odot}\right).
\label{eq:stline_2}
\end{equation}
Galaxies located above \autoref{eq:stline_1} are classified as red, while those lying below \autoref{eq:stline_2} belong to the blue cloud. Galaxies falling between these two relations constitute the green-valley population. 

\subsection{Comparison of Fuzzy-based and empirical methods for physical properties of red, green, and blue galaxies}

We examine the distributions of stellar mass, specific star formation rate (sSFR), and concentration index ($r_{90}/r_{50}$) for galaxies classified using the fuzzy-based method and  the emirical method proposed by \citet{schawinski14}. The results are shown in \autoref{fig:rgb_enq}.

\subsubsection{Red Sequence galaxies}

The red sequence galaxies identified by both methods exhibit similar stellar mass distributions (top left panel of \autoref{fig:rgb_enq}), with nearly identical probability density functions. 
It is important to note that the galaxy sample used in this work is not subject to any explicit stellar-mass threshold. Rather, we construct a volume-limited sample by imposing an absolute-magnitude limit, as shown in the top panel of the \autoref{fig:volume_limited}. This selection restricts the sample to galaxies within a well defined luminosity range. Given the correlation between stellar mass and luminosity, the adopted magnitude limit preferentially excludes galaxies with lower stellar masses, leading to a reduced representation of the low-mass end of the stellar-mass distribution.

Their is a clear differences emerge in their star formation and structural properties. The sSFR distribution for empirically classified red galaxies shows a secondary peak at higher sSFR values (middle left panel of \autoref{fig:rgb_enq}), indicating contamination by star-forming systems. In contrast, the fuzzy-classified red galaxies display a single, well-defined peak at low sSFR, consistent with a uniformly quenched population. This difference is further supported by the concentration index distribution. While both methods predominantly identify bulge-dominated systems ($r_{90}/r_{50} > 2.6$), the empirical classification includes a larger fraction of disc-dominated galaxies (top right panel of \autoref{fig:rgb_enq}). These systems are likely responsible for the elevated sSFR tail, highlighting the limitations of hard boundary-based classification.

\subsubsection{Green Valley galaxies}

The green-valley population exhibits the most significant differences between the two methods. As shown in middle right panel of \autoref{fig:rgb_enq}, the empirical classification yields a higher fraction of disc-dominated galaxies ($r_{90}/r_{50} < 2.6$), whereas the fuzzy classification preferentially selects systems with higher concentration indices, indicative of more evolved morphologies. The sSFR distributions shown in the middle middle panel \autoref{fig:rgb_enq} further reinforce this distinction. Although both samples peak near $\sim 10^{-2}\,\mathrm{Gyr}^{-1}$, the fuzzy-classified green-valley galaxies extend more strongly toward lower sSFR values. In contrast, the empirical sample shows an excess at higher sSFR, suggesting contamination from blue cloud galaxies. The middle left panel of \autoref{fig:rgb_enq} shows that the stellar mass distribution of green valley galaxies for the two methods peak in the mass range $10^{10.75}\,M_\odot$ to $10^{10.9}\,M_\odot$ and look quite similar. At the high mass end (beyond $10^{11.1}\,M_\odot$), the two distributions exhibit significant overlap. These results indicate that the fuzzy classification isolates a population that is more consistently transitional, both in terms of star formation activity and structural evolution. The inclusion of disc-dominated, actively star-forming galaxies in the empirical sample suggests that hard cuts in parameter space are insufficient to cleanly separate the green valley.

\subsubsection{Blue Cloud galaxies}

For blue cloud galaxies, both classification methods (the three bottom panels of \autoref{fig:rgb_enq}) yield broadly consistent results. The majority of galaxies exhibit low concentration indices ($r_{90}/r_{50} < 2.6$), confirming their disc-dominated morphology. Similarly, the sSFR distributions indicate actively star-forming populations in both cases. Minor differences are observed in the stellar mass distributions, with the fuzzy classification showing a slightly higher amplitude at the high-mass end. However, these variations are small compared to the differences observed in the red and green populations. Overall, the agreement between the two methods for blue galaxies suggests that contamination is less significant in this regime, and that both approaches reliably identify actively star-forming systems.

\subsection{AGN Fraction as a function of stellar mass}

We examine the fraction of active galactic nuclei (AGN) as a function of stellar mass for red, green, and blue galaxies identified using our method and the empirical method, as shown in \autoref{fig:rgb_agn}. The motivation for this comparison stems from observational studies suggesting that AGNs may play a crucial role in quenching star formation in green valley galaxies \citep{nandra07, cimatti13, zhang21}. We find that the AGN fraction increases with stellar mass across all galaxy populations, consistent with previous studies. However, at fixed stellar mass, there is no statistically significant difference in the AGN fraction between the two methods for all three galaxy populations. 

\subsection{Clustering properties of galaxy populations}

To further assess the physical consistency of the classification, we compute the two-point correlation function $\xi(r)$ using the Landy--Szalay estimator. The Landy \& Szalay estimator \citep{landy93} is given by,
\begin{equation}
    \xi(r) = \frac{DD(r) - 2DR(r)+ RR(r)}{RR(r)}
    \label{eq:2pcf_estimator}
\end{equation}
where $DD(r)$, $RR(r)$, and $DR(r)$ are normalise counts for data-data, random-random, and data-random pairs at separation $r$.

\begin{figure*}[h!]
\centering
    \includegraphics[width = \textwidth]{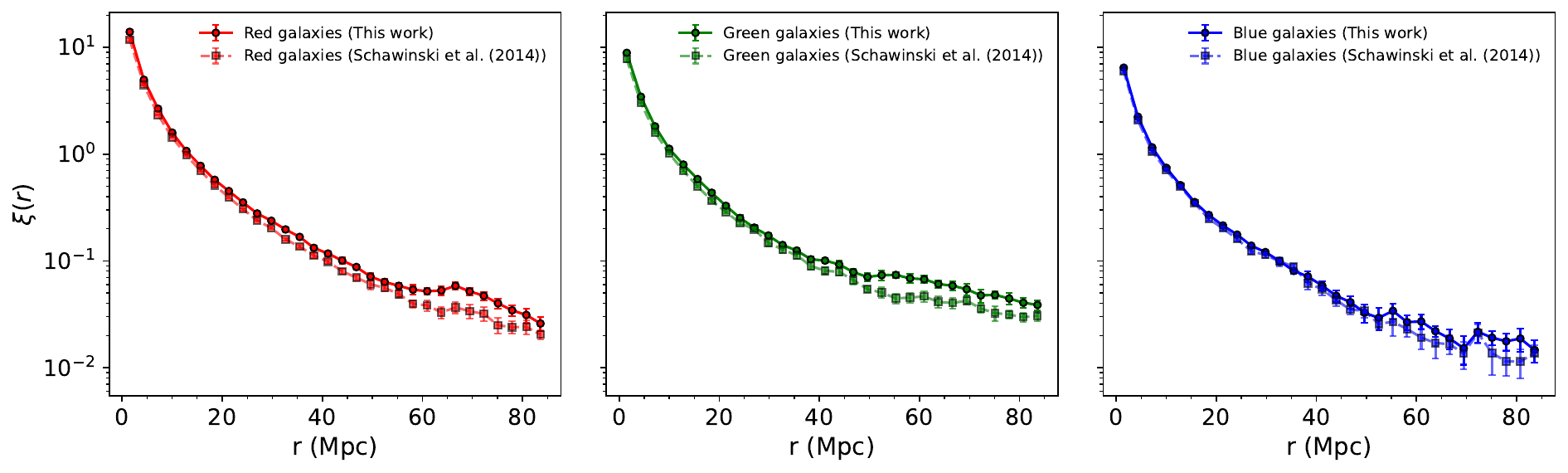}
    \caption{Two-point correlation function, $\xi(r)$, as a function of separation $r$ for red-sequence, green-valley, and blue-cloud galaxies, using our fuzzy based method and for the galaxies classified using the method proposed by \citet{schawinski14}.}
 \label{fig:2pcf_comp}
\end{figure*}

We compare the results obtained using the galaxy classification scheme proposed by \citet{schawinski14} with those derived from our fuzzy-based classification method. The results are shown in \autoref{fig:2pcf_comp}. In both classification schemes, the correlation function decreases monotonically with increasing separation, reflecting the expected transition from strong small-scale clustering to weaker large-scale correlations. The measurements are well described by a power-law model
\begin{equation}
\xi(r) = \left(\frac{r}{r_0}\right)^{-\gamma},
\end{equation}
with best-fit parameters listed in \autoref{tab:2pcf_fits_comp}.

For the fuzzy-classified galaxies, the clustering amplitude follows the expected hierarchy: red galaxies are the most strongly clustered, followed by green-valley and blue cloud galaxies. This trend is consistent with the established connection between galaxy colour, environment, and halo mass. While both classification methods yield broadly similar clustering parameters, notable differences emerge on large scales. 
In particular, the fuzzy-classified red and green-valley galaxies exhibit enhanced clustering signals over the ranges $60$--$80\,\mathrm{Mpc}$ and $50$--$80\,\mathrm{Mpc}$, respectively, compared to the empirical sample. This enhancement likely reflects differences in the halo populations traced by the two methods. The fuzzy classification appears to preferentially select galaxies residing in more massive and highly biased dark matter halos, consistent with their lower sSFR and higher concentration indices. These results further support the conclusion that the fuzzy classification provides a more physically meaningful separation of galaxy populations, particularly in regimes where traditional methods suffer from contamination.

\begin{table}
\centering
\resizebox{\columnwidth}{!}{
\begin{tabular}{lcccc}
\hline
& \multicolumn{2}{c}{\textbf{This work}} & \multicolumn{2}{c}{\textbf{\cite{schawinski14}}} \\
Galaxy & $r_0$ & $\gamma$ & $r_0$ & $\gamma$ \\
\hline
Red   & $13.4360 \pm 0.0901$ & $1.3307 \pm 0.0131$ & $12.5892 \pm 0.0522$ & $1.3575 \pm 0.0084$ \\
Green & $11.0258 \pm 0.0750$ & $1.2378 \pm 0.0108$ & $9.7878 \pm 0.0632$ & $1.3087 \pm 0.0150$ \\
Blue  & $7.6855 \pm 0.0480$ & $1.2779 \pm 0.0130$ & $7.1479 \pm 0.0615$ & $1.2502 \pm 0.0144$ \\
\hline
\end{tabular}
}
\caption{Best-fit correlation length ($r_0$) and slope ($\gamma$) for red, green, and blue galaxies. Results are shown for both the fuzzy-based and \citet{schawinski14} classification methods.}
\label{tab:2pcf_fits_comp}
\end{table}


\section{Conclusion}
\label{sec:conclusion}

We have developed a multi-parameter fuzzy set-based framework for classifying galaxies into the red sequence, green valley, and blue cloud, offering a physically motivated and flexible alternative to traditional hard-cut methods. By incorporating multiple observables such as $(u-r)$ colour, $\log_{10}\left(\frac{\mathrm{sSFR}}{\mathrm{Gyr}^{-1}}\right)$, and $D4000$ and assigning continuous membership values, our approach naturally captures the intrinsic continuity of galaxy properties and provides a more realistic representation of transitional systems. Applied to a volume-limited sample from SDSS, this framework yields a robust and contamination-resistant classification, while remaining easily extensible to additional bimodal observables.

A direct comparison with the empirical classification scheme of \citet{schawinski14} reveals clear and systematic differences. While both methods produce broadly consistent results for blue cloud galaxies, significant discrepancies arise in the red sequence and green valley populations. The empirical approach introduces clear contamination from star-forming and disc-dominated systems, reflecting the limitations of rigid boundaries in colour--stellar mass space. In contrast, the fuzzy classification isolates galaxy populations that are more physically coherent, exhibiting cleaner separations in star formation activity and structural properties. This leads to more consistent and interpretable distributions across key diagnostics, including sSFR and morphology.

Despite these differences, the AGN fraction as a function of stellar mass remains statistically consistent between the two methods for all three galaxy populations. This suggests that both approaches capture the global trends of AGN activity, although the improved sample purity in the fuzzy classification provides a more reliable basis for physical interpretation.

Further insights emerge from the clustering analysis. While the overall clustering trends are similar, subtle but significant differences are observed on large scales. In particular, the enhanced clustering amplitude of fuzzy-classified red and green-valley galaxies indicates that our method preferentially selects systems residing in more strongly biased dark matter halos. These differences highlight the sensitivity of large-scale structure measurements to classification methodology, especially through the impact of misclassified galaxies on the inferred halo occupation and the two-halo term.

Taken together, our results demonstrate that the proposed fuzzy set-based framework provides a more robust, physically consistent, and data-driven classification of galaxy populations. By avoiding arbitrary boundaries and embracing the continuous nature of galaxy evolution, this approach enables a clearer and more reliable interpretation of galaxy properties and their underlying physical processes.

We also find that the main conclusions of our analysis remain unchanged when alternative GMM fitting strategies, including peak-based initialization and constrained component means, are adopted. This further demonstrates the robustness of our classification framework against variations in the GMM fitting procedure.

The framework presented here can be extended in several important directions. Future studies may incorporate additional galaxy observables to further refine the classification and capture a broader range of physical processes. In addition, applying this approach to deeper and higher-redshift surveys such as LSST and Euclid will enable the investigation of galaxy evolution across cosmic time.

Finally, we emphasize that classification choices play a crucial role in shaping our understanding of galaxy evolution. Methods based on fixed empirical boundaries can lead to biased or inconsistent interpretations, particularly in transitional regimes such as the green valley. In this context, data-driven approaches like the one presented here are essential for advancing a more accurate and unified picture of galaxy evolution.

\section*{Acknowledgements}
AM thanks Anindita Nandi for help with the SDSS data. AM acknowledges UGC, Government of India for support through a Junior Research Fellowship. BP acknowledges IUCAA, Pune, for providing support through the associateship programme. BP acknowledges financial support from Government of India through the project ANRF/ARG/2025/000535/PS. 

Funding for the SDSS and SDSS-II has been provided by the Alfred
P. Sloan Foundation, the Participating Institutions, the National
Science Foundation, the U.S. Department of Energy, the National
Aeronautics and Space Administration, the Japanese Monbukagakusho, the
Max Planck Society, and the Higher Education Funding Council for
England. The SDSS website is http://www.sdss.org/.

The SDSS is managed by the Astrophysical Research Consortium for the
Participating Institutions. The Participating Institutions are the
American Museum of Natural History, Astrophysical Institute Potsdam,
University of Basel, University of Cambridge, Case Western Reserve
University, University of Chicago, Drexel University, Fermilab, the
Institute for Advanced Study, the Japan Participation Group, Johns
Hopkins University, the Joint Institute for Nuclear Astrophysics, the
Kavli Institute for Particle Astrophysics and Cosmology, the Korean
Scientist Group, the Chinese Academy of Sciences (LAMOST), Los Alamos
National Laboratory, the Max-Planck-Institute for Astronomy (MPIA),
the Max-Planck-Institute for Astrophysics (MPA), New Mexico State
University, Ohio State University, University of Pittsburgh,
University of Portsmouth, Princeton University, the United States
Naval Observatory, and the University of Washington. 

\bibliography{ms}

\appendix

\section{Sample selection and visual comparison of galaxy classifications}
\label{app:one}

\begin{figure*}
    \centering
    \includegraphics[width=0.7\textwidth]{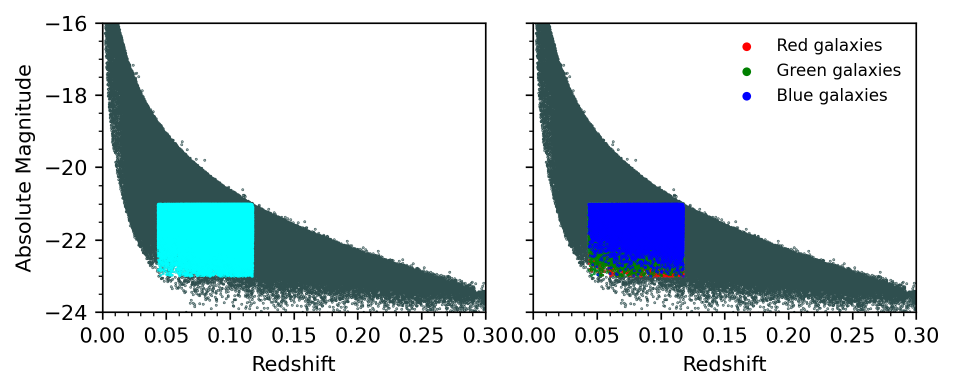}
    \caption{The left panel of this figure shows the distribution of SDSS DR18 galaxies in the redshift--absolute magnitude plane. The cyan shaded region represents our volume-limited sample. The right panel shows the galaxies within the final sample, colour-coded according to their blue, green, and red classifications, illustrating the distribution of the three populations within the volume-limited selection region.}
    \label{fig:volume_limited}
\end{figure*}

In this Appendix, we first demonstrate the selection of our volume-limited galaxy sample using the redshift--absolute magnitude distribution of the SDSS DR18 galaxies shown in \autoref{fig:volume_limited}. The volume-limited selection region is indicated by the cyan shaded region in the left panel, while the right panel shows the distribution of the final blue, green, and red galaxy populations within this region. This provides a visual assessment of the representation of the different galaxy populations in the volume-limited sample.

\begin{figure*}
    \centering
    \includegraphics[width=\textwidth]{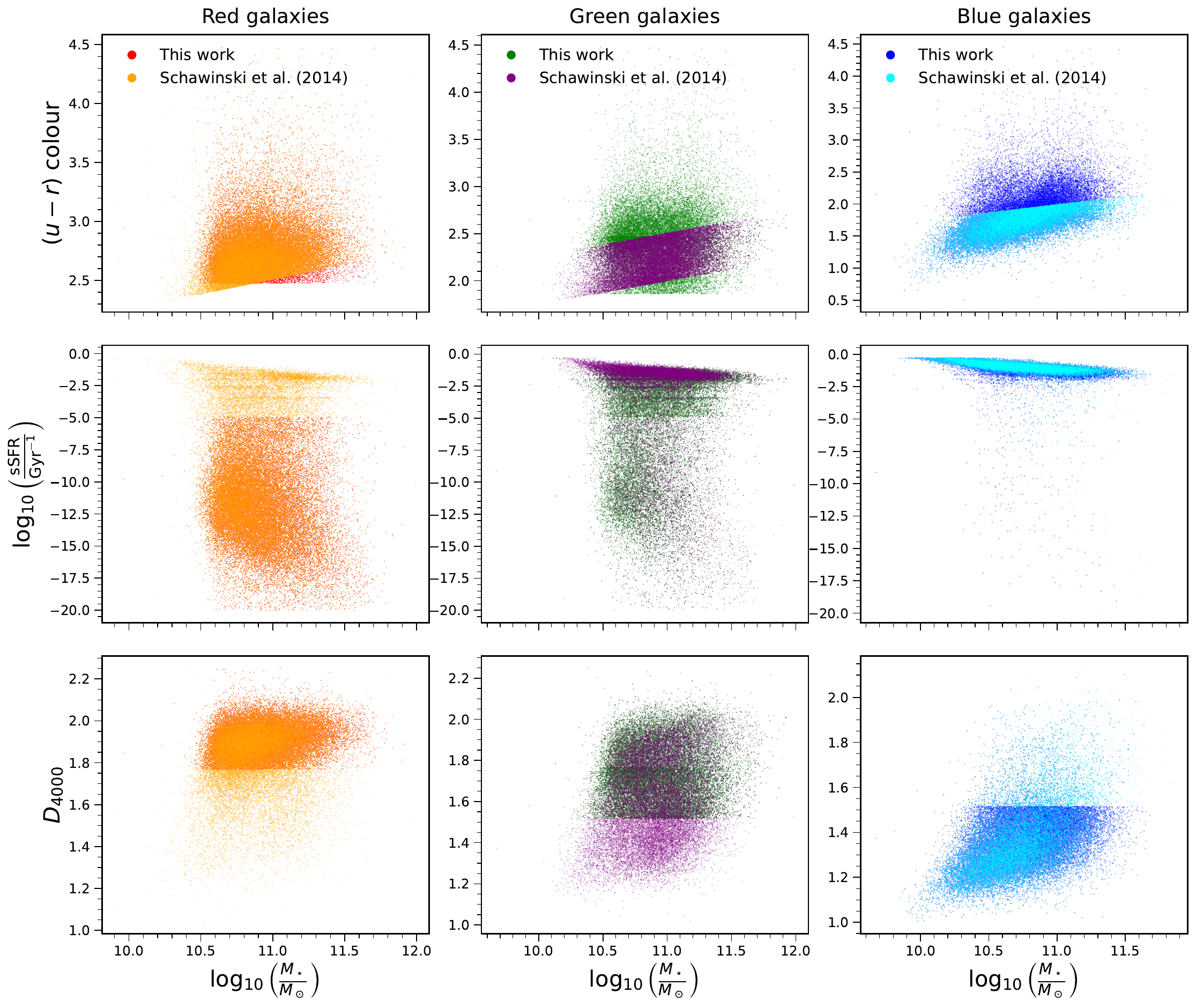}
    \caption{Comparison of galaxy classifications in the stellar mass--$(u-r)$ colour, stellar mass--sSFR, and stellar mass--$D4000$ planes. Points are colour-coded according to their respective classifications in this work and those of \citet{schawinski14}.}
    \label{fig:classification_appendix}
\end{figure*}

We subsequently compare the different galaxy classifications by examining the distributions of galaxies in the stellar mass--$(u-r)$ colour, stellar mass--sSFR, and stellar mass--$D_{4000}$ planes in \autoref{fig:classification_appendix}.


\end{document}